\documentclass[iop]{emulateapj}

\slugcomment{{\sc Accepted for publication in ApJ:} April 9, 2012}

\usepackage{multirow}
\usepackage{hyperref}

\newcommand{\lsim}{\lower 2pt \hbox{$\, \buildrel {\scriptstyle
<}\over {\scriptstyle \sim}\,$}}  \newcommand{\gsim}{\lower 2pt
\hbox{$\, \buildrel {\scriptstyle >}\over {\scriptstyle \sim}\,$}}

\newcommand{\Pa}{Pa${\alpha}$}

\newcommand{\Brg}{Br${\gamma}$}

\begin{document}


\title{Spatially resolved kinematics of the central regions of M83: hidden mass signatures and the role of supernovae}
\author{J. Piqueras L\'opez\altaffilmark{1},
R. Davies\altaffilmark{2},
L. Colina\altaffilmark{1},
G. Orban de Xivry\altaffilmark{2},
}
\affil{$^1$ Centro de Astrobiolog\'ia, INTA-CSIC, Spain\\
$^2$ Max-Planck-Institut f\"ur extraterrestrische Physik, Postfach 1312, 85741, Garching, Germany}
\email{piqueraslj@cab.inta-csic.es}

\begin{abstract}
The barred grand-design spiral M83 (NGC~5236) is one of the most studied galaxies given its proximity, orientation, and particular complexity. Nonetheless, many aspects of the central regions remain controversial conveying our limited understanding of the inner gas and stellar kinematics, and ultimately of the nucleus evolution.

In this work, we present AO VLT-SINFONI data of its central $\sim235\times140$\,pc with an unprecedented spatial resolution of $\sim$0.2\,arcsec, corresponding to $\sim$4\,pc. We have focused our study on the distribution and kinematics of the stars and the ionised and molecular gas by studying in detail the \Pa\ and \Brg\ emission,  the H$_2$ 1-0S(1) line at 2.122\,$\mu$m and the [FeII] line at 1.644\,$\mu$m, together with the CO absorption bands at 2.293\,$\mu$m and 2.323\,$\mu$m. Our results reveal a complex situation where the gas and stellar kinematics are totally unrelated. Supernova explosions play an important role in shaping the gas kinematics, dominated by shocks and inflows at scales of tens of parsecs that make them unsuitable to derive general dynamical properties.

We propose that the location of the nucleus of M83 is unlikely to be related to the off-centre `optical nucleus'. The study of the stellar kinematics reveals that the optical nucleus is a gravitationally bound massive star cluster with $M_{\rm dyn} = (1.1\pm0.4)\times10^7\,M_{\odot}$, formed by a past starburst. The kinematic and photometric analysis of the cluster yield that the stellar content of the cluster is well described by an intermediate age population of $\log T(\rm yr) = 8.0\pm0.4$, with a mass of $M^{\rm \star} \simeq (7.8\pm2.4)\times10^6\,M_{\odot}$.
\end{abstract}

\keywords{galaxies:individual (M83, NGC~5236) -- galaxies:general -- galaxies:starburst -- galaxies:nuclei -- galaxies:kinematics and dynamics -- galaxies:structure}

\section{Introduction}

M83 (NGC~5236) is a nearby ($D=4.6$\,Mpc, 22\,pc\,arcsec$^{-1}$, $z=0.0017$ from the NASA/IPAC Extragalactic Database, NED) barred grand-design spiral galaxy with a nuclear starburst. The galaxy has been object of intense study during the last decade, given the complexity of its central regions, its proximity and the fact that it is almost face-on, with an inclination of $i=24^{\circ}$ \citep{Comte:1981p6710}. This makes it a good candidate on which to make use of high spatial resolution integral field spectroscopy (IFS) to study the controversial aspects of its innermost regions.

The general morphology of the galaxy shows a pronounced bar and well-defined spiral arms where star formation is intense. On the other hand, the central regions of M83 in the infrared are rather complex. The J--K images of the inner region show two non-concentric circumnuclear dust rings, which are associated with two inner Lindblad resonances \citep{Elmegreen:1998p5640}. These two rings are connected by an inner bar, almost perpendicular to the main stellar bar. The general shape of the extended emission traces an arc  between these two dust rings, where the star formation is concentrated.

The location of the nucleus of M83 remains unclear. \cite{Thatte:2000p4743} first reported the existence of a 3.4\,arcsec ($\sim75$\,pc) offset between the optical nucleus and the centre of symmetry of the bulge K-band isophotes. The centre of symmetry of these external isophotes is coincident with the  dynamical centre proposed by \cite{Sakamoto:2004p5705}, based on two-dimensional CO spectroscopy, and confirmed later by \cite{Rodrigues:2009p5531} and \cite{Knapen:2010p5390}. Different locations of hidden mass concentrations were proposed to host the supermassive black hole of M83, mainly based on studies of the gas kinematics (\citealt{Mast:2006p4849}, \citealt{Diaz:2006p4776}, \citealt{Rodrigues:2009p5531} and \citealt{Knapen:2010p5390}).

In this paper, we present new integral field VLT-SINFONI \citep{Eisenhauer:2003p8484} spectroscopy in H+K bands, with an unprecedented spatial resolution, covering the central $\sim235\times140$\,pc of the galaxy. We study the stellar and gas kinematics of the inner parts and address some of the open questions regarding the hidden mass concentrations at off-nuclear locations (\citealt{Thatte:2000p4743}, \citealt{Mast:2006p4849}, \citealt{Diaz:2006p4776}) and the origin of the steep velocity gradients in the gas kinematics \citep{Rodrigues:2009p5531}, revealing a complex scenario where supernovae play a key role in the kinematics.

\section{Observations}
\subsection{Observations and Data Reduction}

\begin{figure}[t]
\begin{center}
\includegraphics[angle=90, width=0.48\textwidth]{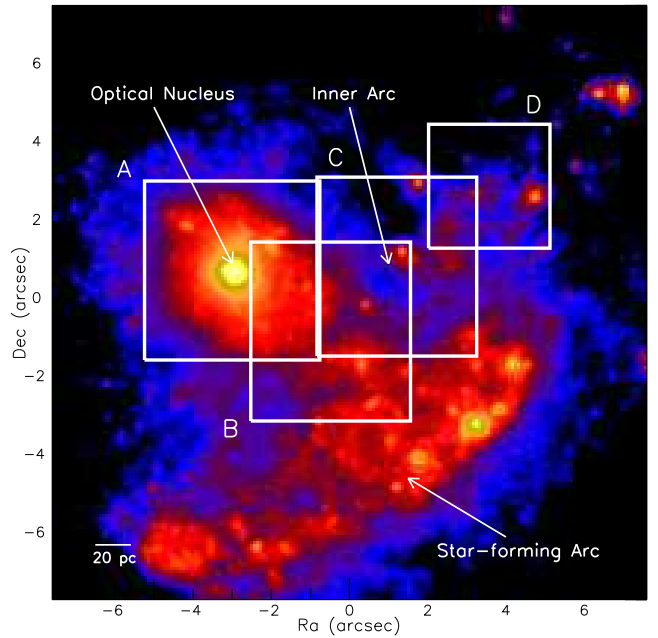}
\caption{Central region of M83: HST NICMOS F222M image with the fields covered by the four SINFONI pointings superimposed. The location of the optical nucleus and the arc of star formation mentioned in the text are also shown. Although the inner arc is not visible in the continuum map, it is clearly visible in the western part of the FoV in the flux panels in Fig.~\ref{figure:maps}. The total coverage of our SINFONI observations is $\sim235\times140$\,pc.}
\label{figure:pointings}
\end{center}
\end{figure}

The M83 observations are divided into four different pointings, labeled as A, B, C and D in Fig.~\ref{figure:pointings}. These pointings were chosen to cover the stellar nucleus of the galaxy, that is identified with the optical nucleus of M83, covered by pointing A; the centre of symmetry of the bulge K-band isophotes \citep{Thatte:2000p4743}, that corresponds to pointing B;  the proposed location of a hidden mass by \cite{Mast:2006p4849}, by pointing C; and the putative massive black hole location given in \cite{Diaz:2006p4776} sampled by pointing D. The first three, A, B and C, were carried out between April and June 2009 in service mode, using the AO module fed by a LGS. The fourth was performed in July 2011, also in service mode. The data were taken in the H+K configuration, using a scale plate of $0.05\times0.1$\,arcsec\,pixel$^{-1}$ that yields a nominal field of view (FoV) of $\sim$3.2" $\times$ 3.2" then enlarged by dithering. The wavelength range covered is from 1.45\,$\mu$m to 2.46\,$\mu$m with a spectral resolution of R$\sim$1500.

Although a total of four pointings were programmed, the main analysis in this work has been performed on three of them, A, B and C. The observations for pointing D could not be completed and there was only 1 usable object frame. Taking into account the vast difference of quality in the data, we decided not to include the new data in the main analysis, but use it instead to support some of the results. The first three pointings cover an area of $\sim$8" $\times$ 6" around the nucleus of the galaxy, while the fourth pointing covers $\sim3"\times3"$. The footprints of the observed pointings are shown in Fig.~\ref{figure:pointings}. The total integration time was 3300\,s for pointing A and 3600\,s for each of pointings B and C, split into individual exposures of 300\,s. In addition, four sky frames of 300\,s were taken for every pointing every two on-source exposures to subtract the sky emission, following the pattern OOSOOSOO. In the same way, a total of five standard stars (Hip066957, Hip069230, Hip070506, Hip071136 and Hip098641) were observed in order to perform telluric and flux calibration. The fourth pointing is a single on-source exposure of 300\,s with a matching sky frame. The standard star used for the calibration was Hip001115.

The reduction of the data was performed using the standard ESO pipeline. An individual cube was built from each frame, from which the background sky emission was subtracted using the method outlined in \cite{Davies:2007p2525}. We performed the telluric and flux calibration on each cube individually to improve the results. Taking into account the relative shifts of the jitter pattern, we then combined the data to create a single cube for each pointing. Finally, these were combined to build a single mosaic.

The telluric and flux calibration were performed in two steps. First, each individual star was normalised to the continuum level, using a blackbody profile at the T$_{\rm eff}$ listed in the Tycho-2 Spectral Type Catalog \citep{Wright:2003p4322}. To remove the absorption features in the spectra of the stars, we used a solar template, convolved and binned to match the resolution of the SINFONI data. The result is a ``sensitivity function" that takes into account the atmospheric transmission.
Secondly, we used the H and K magnitudes of the stars from the 2MASS catalog \citep{Skrutskie:2006p5780} to convert our spectra from counts to physical units. We made use of the response curves of 2MASS filters, as defined in \cite{Cohen:2003p3372}, to obtain the values in counts of our spectra at their effective wavelengths. Using the above mentioned magnitudes, we obtained two conversion factors for each star (one for H-band and one for K-band). These factors were almost identical for both bands in every star, which verifies the calibration. We adopted the mean value to scale our curves. The flux-calibrated cubes were obtained by dividing the individual cubes by the ``sensitivity function", to correct from the atmospheric transmission, and by multiplying them by the conversion factor. 

\subsection{Gas and Stellar Kinematics}

The gas kinematics were extracted by fitting a gaussian profile to the most relevant emission lines, using the code \emph{LINEFIT} described in \cite{Davies:2011ApJ741} (see also \citealt{ForsterSchreiber:2009p7378}). During the extraction, the OH sky line at 2.18\,$\mu$m is used to remove the instrumental broadening, measured to be 13\,\AA\ FWHM.

\begin{figure}[t]
\begin{center}
\includegraphics[angle=180, width=0.48\textwidth]{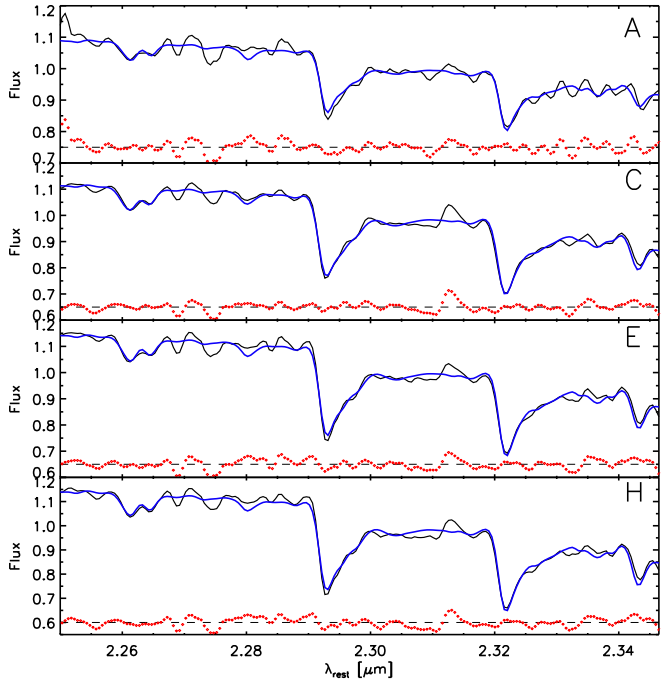}
\caption{Integrated spectra of apertures A, C, E and H, represented in Fig.~\ref{figure:maps}) and the result from fitting template spectra to the stellar absorption features. The normalised spectra are shown in black, and the results from the fitting are plotted as a thick blue line. The red dotted lines show the residuals from subtracting the fit.}
\label{figure:ppxf}
\end{center}
\end{figure}

To extract the stellar kinematics, we focused on the two most prominent CO bands, CO (2 -- 0) at 2.293\,$\mu$m and CO (3 -- 1) at 2.323\,$\mu$m, and used the Penalized Pixel-Fitting (pPXF) software developed by \cite{Cappellari:2004p4916} to fit a library of stellar templates to our data (see Fig.~\ref{figure:ppxf}). We made use of the near-IR library of spectral templates from \cite{Winge:2009p4732}, which covers the wavelength range of 2.15\,$\mu$m -- 2.43$\mu$m with a spectral resolution of $R\sim5600$ and sampled at 1\,\AA\,pixel$^{-1}$. The library contains a total of 23 late-type stars, from F7III to M3III, and was previously convolved to our SINFONI resolution.

The uncertainty in the gas and stellar kinematics is highly dependant on the S/N and how well resolved the line is. For the gas kinematics, the uncertainties could range from $\sim1$\% of the resolution element in those regions with high S/N up to more than $\sim$10\% in the regions with poorer S/N. The kinematic precision achievable with the stellar absorption features also depends on the S/N, although it is typically lower than the precision in the gas kinematics mainly due to uncertainties in the template matching uncertainties. As shown in Fig.~\ref{figure:ppxf}, the quality of the template fitting indicates that the precision we achieved in the stellar kinematics is high, less than $\sim$40\,km\,s$^{-1}$ in those regions with high S/N, and that offsets of  $\gsim$30\,km\,s$^{-1}$ would be clearly visible in the residuals of the fitting.

In order to compare the velocity fields of the different phases of the gas and the stellar component, we have established a reference value of $cz = 589.6$\,km\,s$^{-1}$ for the velocity that has been used as a zero-point for the velocity maps. This reference value for the velocity has been chosen as the mean value of the stellar velocity in a small aperture of 5 spaxel radius centred in the dynamical centre of the galaxy proposed by \cite{Sakamoto:2004p5705}.  

\subsection{Voronoi Binning}

Before the extraction of the kinematics, the data were binned to achieve a minimum value of S/N on the whole field. We used the Voronoi binning method implemented by \cite{Cappellari:2003p4908} to maintain the maximum spatial resolution of our maps while constraining the minimum S/N ratio.

The Voronoi tessellation employs bins of approximate circular shape to divide the space, described in terms of a set of points called \emph{generators}. Every bin encloses all the points that are closer to its generator than any other generator. The algorithm finds an initial set of generators by selecting the spaxel with highest S/N ratio and accreting spaxels to that bin, until the required threshold is reached. Then, it moves downwards to lower S/N spaxels until all the points are assigned to a proper bin. This set of generators is then refined to satisfy both topological and morphological criteria and to ensure that the scatter of the S/N among all the bins is reduced to a minimum (see \citealt{Cappellari:2003p4908} for further details).


\begin{deluxetable}{cc}
\tablecolumns{2}
\tablewidth{100pt}
\tablecaption{S/N thresholds used for the Voronoi binning\label{table:S/N}}
\tablehead{\colhead{Feature} & \colhead{S/N} \\}
\startdata
\Pa & 50\\
\Brg & 45\\
\ H$_{2}$ 1-0S(1) & 15\\
\ [FeII] & 25\\
\ Stellar & 50\\
\enddata
\end{deluxetable}

This binning of the data does not affect those spaxels with high S/N ratio, preserving the original spatial resolution of these regions. The maps obtained for each line and the continuum are then binned independently, since their flux distributions are totally different. We have therefore defined different S/N thresholds for each line, in order to obtain appropriately sampled maps. The S/N cutoff used for each line and for the stellar continuum are shown in Table~\ref{table:S/N}. We selected these values to achieve roughly the same number of bins in each map.

\subsection{PSF Determination}

The measurement of the PSF has been done by comparing the SINFONI data to higher resolution HST NICMOS data. As discussed in \cite{Davies:2008p5959}, we can use a higher resolution image with a well known PSF to estimate the PSF of a lower resolution image. After resampling the NICMOS image to our SINFONI pixel scale, the aim is to find a broadening function, $B$, that satisfies $I_{SINFONI} = I_{NICMOS}\otimes B$. Since the NICMOS PSF is well known, we can estimate the SINFONI PSF as $PSF_{SINFONI} = PSF_{NICMOS}\otimes B$. The shape of the resulting PSF is dominated by the broadening function.

We performed independent fittings of the broadening function for each pointing and each band. The NICMOS images were obtained with the NIC2 camera, using the F160W and F222M filters for H and K band respectively, with a pixel scale of 0.075\,arcsec\,pixel$^{-1}$. After trying different models for the broadening function $B$, we find that it is better described as a symmetric double Gaussian with a narrow component of $\sim$1\,pixel FWHM and another wide but fainter component of $\sim$4--6\,pixel FWHM that takes into account the seeing-limited halo. The different values of the FWHM of the resulting PSF are shown in Table~\ref{table:PSF}.

We note that the resolution of our data is limited by the pixel scale chosen for the observations rather than the LGS-AO performance since, in order to cover a wider FoV ($\sim$3\,arcsec), we chose the 0\farcs05$\times$0\farcs1 pixel scale from the three configurations available for SINFONI.


\begin{deluxetable*}{cccccccc}
\tablecolumns{8}
\tablewidth{300pt}
\tablecaption{FWHM of the PSF for each pointing. The values are measured after convolving the broadening function, $B$, obtained from the fitting with the PSF of NICMOS images.\label{table:PSF}}
\tablehead{
\colhead{Pointing} & 
\multicolumn{3}{c}{H-Band FWHM} && 
\multicolumn{3}{c}{K-Band FWHM} \\
\cline{2-4} \cline{6-8} \\
\colhead{} & \colhead{(pixel)} & \colhead{(\arcsec)} & \colhead{(pc)} && \colhead{(pixel)} & \colhead{(\arcsec)} & \colhead{(pc)}\\
}
\startdata
A & 3.66 & 0.18 & 4.03 && 4.05 & 0.20 & 4.45\\
B & 3.76 & 0.19 & 4.14 && 4.18 & 0.21 & 4.60\\
C & 2.95 & 0.15 & 3.24 && 4.09 & 0.20 & 4.50\\
\enddata
\end{deluxetable*}

\section{Overview of Data}

The inner $\sim$190 $\times$ 130\,pc of M83 are covered by pointing A centered on the optical nucleus of the galaxy, pointing B on the photometric centre, and pointing C on the off-nuclear black hole location proposed by \cite{Mast:2006p4849}. 

\begin{figure*}
\begin{center}
\includegraphics[angle=0, width=1\textwidth]{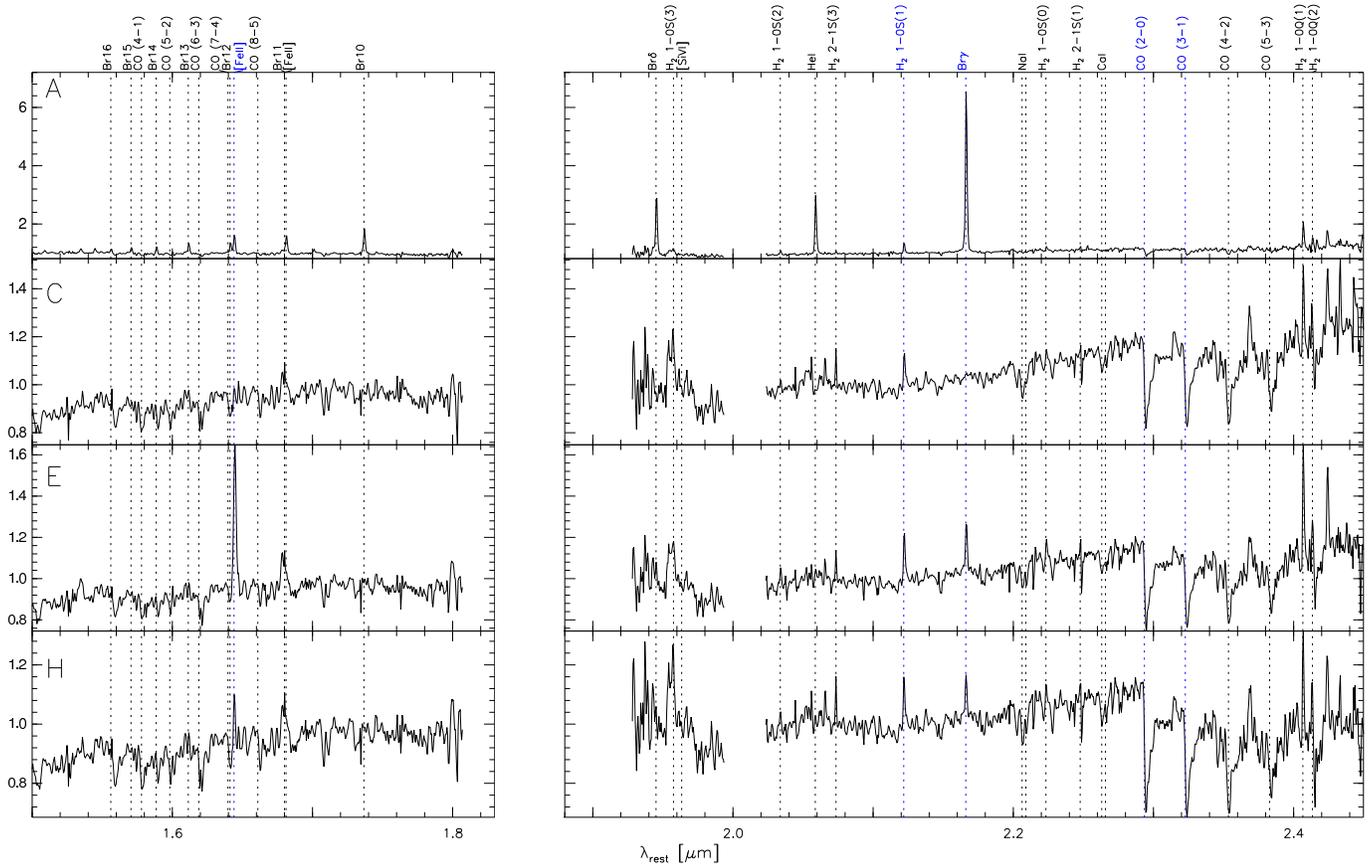}
\caption{Normalised H+K spectra of the apertures A, C, E and H (see Fig.~\ref{figure:maps} for reference). Aperture A is located at the maximum of the \Brg\ emission, aperture C covers the centre of the optical nucleus, aperture E corresponds to one of the bright spots of [FeII] emission next to the optical nucleus and aperture H is located at the position of one of the SNR listed in \cite{Dopita:2010p5365}. The wavelengths of a number of lines and features are identified, and those studied in this paper are identified in blue. These spectra clearly illustrate the wide variety of excitation conditions that are occurring in the inner regions of M83.}
\label{figure:spectra}
\end{center}
\end{figure*}

\begin{figure*}
\centering
\includegraphics[angle=90, width=0.85\textwidth]{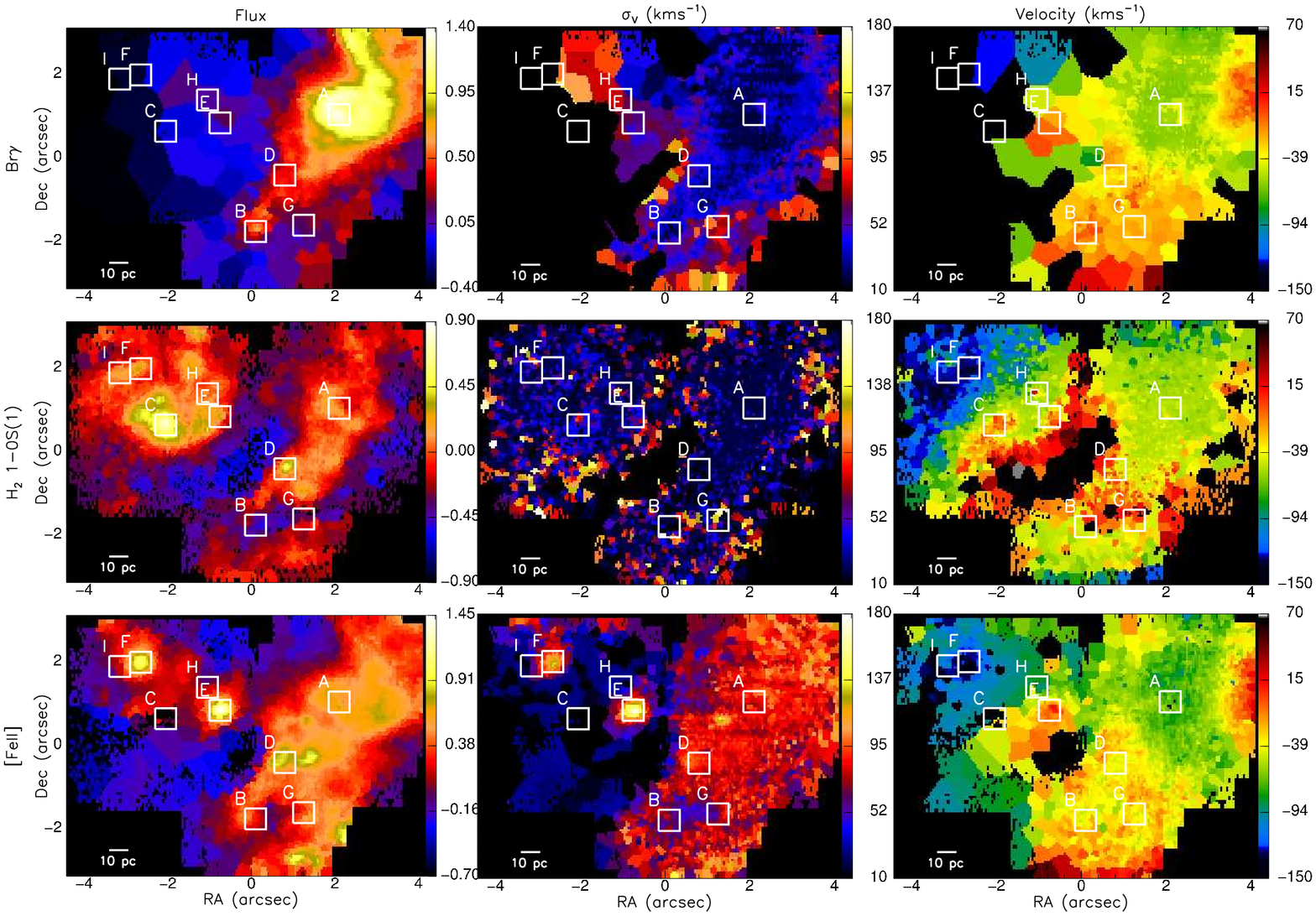}
\caption{Flux and kinematics maps of the main emission lines. From top to bottom, the \Brg\, , H$_{2}$ 1-0S(1)  and [FeII] maps, and from left to right, flux, velocity dispersion and velocity field. The boxes indicate the apertures used to extract spectra, some of which are shown in Fig.~\ref{figure:spectra}. These also act as reference points with respect to the discussion in the text. Flux maps are scaled with a factor 3$\times$10$^{-18}$\,erg\,s$^{-1}$\,cm$^{-2}$.}
\label{figure:maps}
\end{figure*}

The wide spectral coverage of the H+K band configuration of SINFONI allows us to study in detail a large number of emission lines and stellar absorption features (see Fig.~\ref{figure:spectra}).  In order to achieve a good level of S/N in the whole FoV, we focussed our study of the gas kinematics on the brightest emission lines, i.e. \Brg\ 2.166\,$\mu$m for the ionized gas, the ro-vibrational transition H$_{2}$ 1-0S(1) at 2.122\,$\mu$m for the warm molecular gas and the [FeII] line at 1.644\,$\mu$m. We have extracted surface brightness, velocity dispersion and velocity maps of these three lines, represented in Fig.~\ref{figure:maps}, that allow us to study different phases of the interstellar medium. 

Although the \Pa\ line is the brightest in the wavelength range covering the H and K bands, it lies at a wavelength where the atmospheric transmission is very low. Here, the strong variability of the sky absorption makes it very difficult to perform a good correction of the transmission, which translates into an increase of the noise compromising the results of the kinematics extraction. However, given the brightness of the \Pa\ line, we were able to obtain a map of the emission that can be compared with HST/NICMOS archive images of the same region.

As noted above, we make use of the H$_{2}$ 1-0S(1) line to trace the warm molecular gas in the whole FoV. However, the detection of additional H$_{2}$ transitions allows us to study in more detail, in Sec.~\ref{sec:warmh2}, the excitation mechanisms of the H$_{2}$ in the inner regions of M83, and distinguish collisional excitation in shocks from radiative fluorescence. To improve the S/N ratio of the weaker transitions, we have integrated the signal of all the spaxels from the inner star-forming arc (those above a certain flux level) and those from the optical nucleus (see Fig.~\ref{figure:pointings}). This allows us to measure the fluxes of six different transitions with good level of confidence. The different lines measured are listed in Table~\ref{table:h2}. 

\begin{figure*}
\begin{center}
\includegraphics[angle=0, width=1\textwidth]{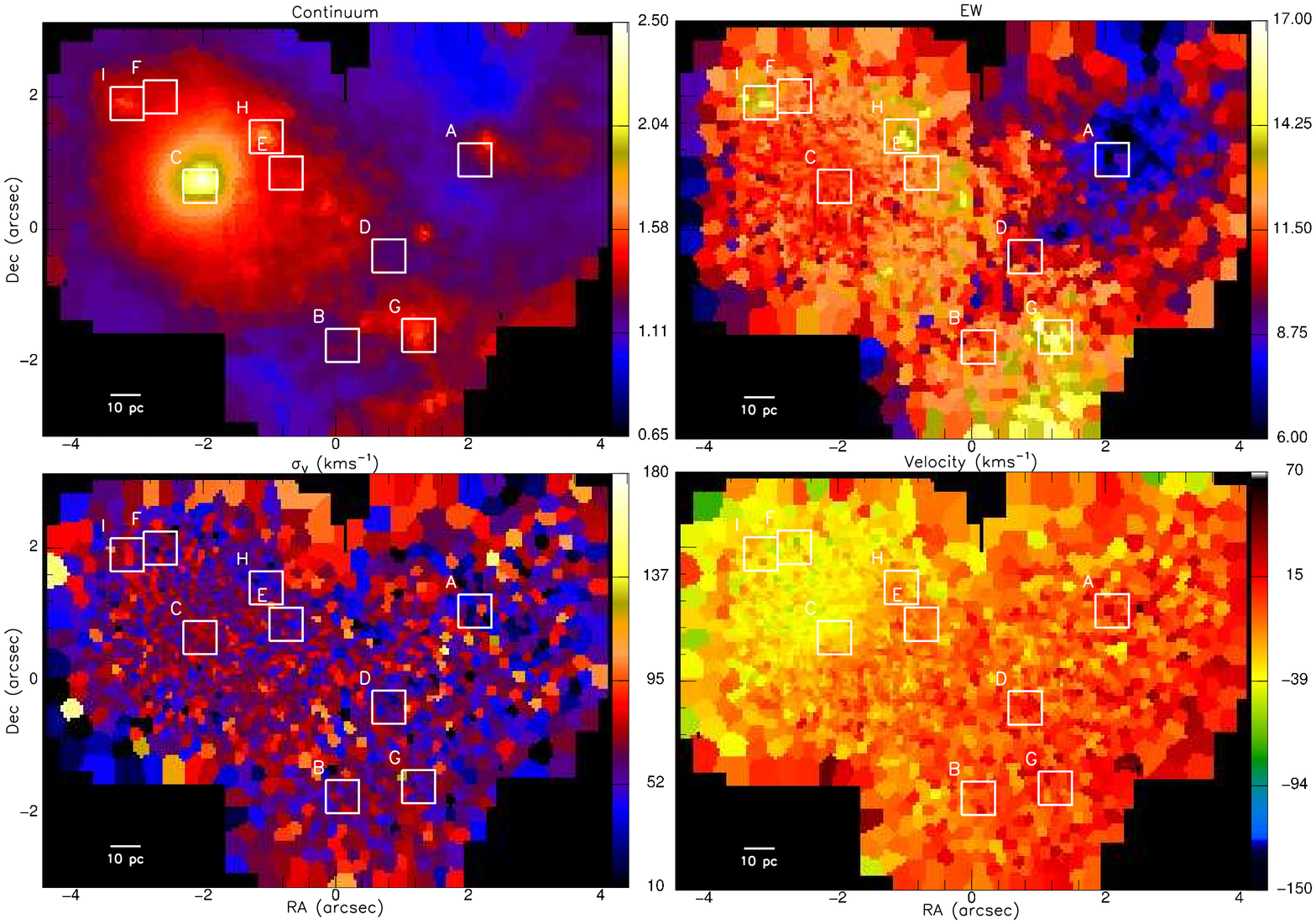}
\caption{Maps of the stellar continuum flux (top left), equivalent width (top right), velocity dispersion (bottom left) and velocity field (bottom right) for the nuclear region of M83. These were obtained by fitting the first two CO absorption bands, CO (2 -- 0) at 2.293\,$\mu$m and CO (3 -- 1) at 2.323\,$\mu$m. The boxes indicate the same apertures as in Fig.~\ref{figure:maps}. Flux map is scaled with a factor 3$\times$10$^{-18}$\,erg\,s$^{-1}$\,cm$^{-2}$.}
\label{figure:stellar}
\end{center}
\end{figure*}

In addition to the numerous emission lines available, we detect a variety of stellar features in our spectra, most notably the NaI, CaI and CO absorption bands. As mentioned previously, we have focussed on the first two CO bands, CO (2 -- 0) at 2.293\,$\mu$m and CO (3 -- 1) at 2.323\,$\mu$m to study the stellar kinematics (Fig.~\ref{figure:stellar}). 

\subsection{Morphology and Kinematics of the Gas}
\label{sec:overview}

As shown in Fig.~\ref{figure:maps}, the morphology and, to some extent, the kinematics of the distinct phases of the gas are rather different. The \Brg\ emission is mainly associated with the inner star-forming arc towards the west side of the FoV. The global velocity gradient of $\sim$60\,kms$^{-1}$ from northwest to southeast is consistent with an inflow of gas along the spiral arms and through the inner bar (\citealt{Elmegreen:1998p5640}, \citealt{Crosthwaite:2002p5369}, \citealt{Fathi:2008p5543}) to the photometric centre of the galaxy. Superimposed on this, there is a ring that is on and around aperture A in Fig.~\ref{figure:maps}. One of the most remarkable properties of this ring-like feature is that it shows no velocity gradient and has a low velocity dispersion that, together, argue against it being a dynamical structure. We discuss this feature in detail in Sec.~\ref{sec:supernovae}.

The H$_{2}$ emission is mainly associated with the inner arc along the western part of the FoV and with the optical nucleus. The kinematics are very similar to those exhibited by the ionized gas. Around region A in Fig.~\ref{figure:maps}, the velocity is the same as that of the \Brg\ emission, where no dominant velocity gradient is observed. Moreover, the emission of the molecular gas resembles the observed ring-like feature of the ionized gas. However, the strong emission of the optical nucleus allowed us to trace the velocity field across this region in more detail. The velocity gradient measured from northeast to southwest is $\sim$200\,km\,s$^{-1}$ in $\sim$45\,pc, significantly steeper than the gradient measured from the stellar kinematics (see Fig.\ref{figure:stellar}). The velocity dispersion shows a similar picture as the \Brg\ emission, where the low values measured along the inner arc suggest that the gas is confined to a thin plane, presumably a disk supported by rotation.

The [FeII] emission is highly extended along the inner arc, showing various knots of strong emission. Two of the brightest spots are located in the outskirts of the optical nucleus, labeled as E and F in Fig.~\ref{figure:maps}, at radial distances of $\sim30$\,pc and $\sim32$\,pc respectively. These two knots also exhibit a high velocity dispersion ($\sim150$\,km\,s$^{-1}$ in region E and $\sim115$\,km\,s$^{-1}$ in region F, both taking into account the spatial resolution of our data) and are probably tracing individual supernovae. We return to this issue in Sec.~\ref{sec:supernovae}. Although the velocity field of the [FeII] is very similar to those traced by the \Brg\ and H$_2$ emission, the velocity dispersion is systematically higher along the inner star-forming arc. As we discuss in Sec.~\ref{sec:supernovae}, this higher velocity dispersion may be a sign of recent supernovae explosions, and would set a constraint on the age of the stellar populations along the arc.

\subsection{Stellar Component}

\begin{figure}[t]
\begin{center}
\includegraphics[angle=0, width=0.48\textwidth]{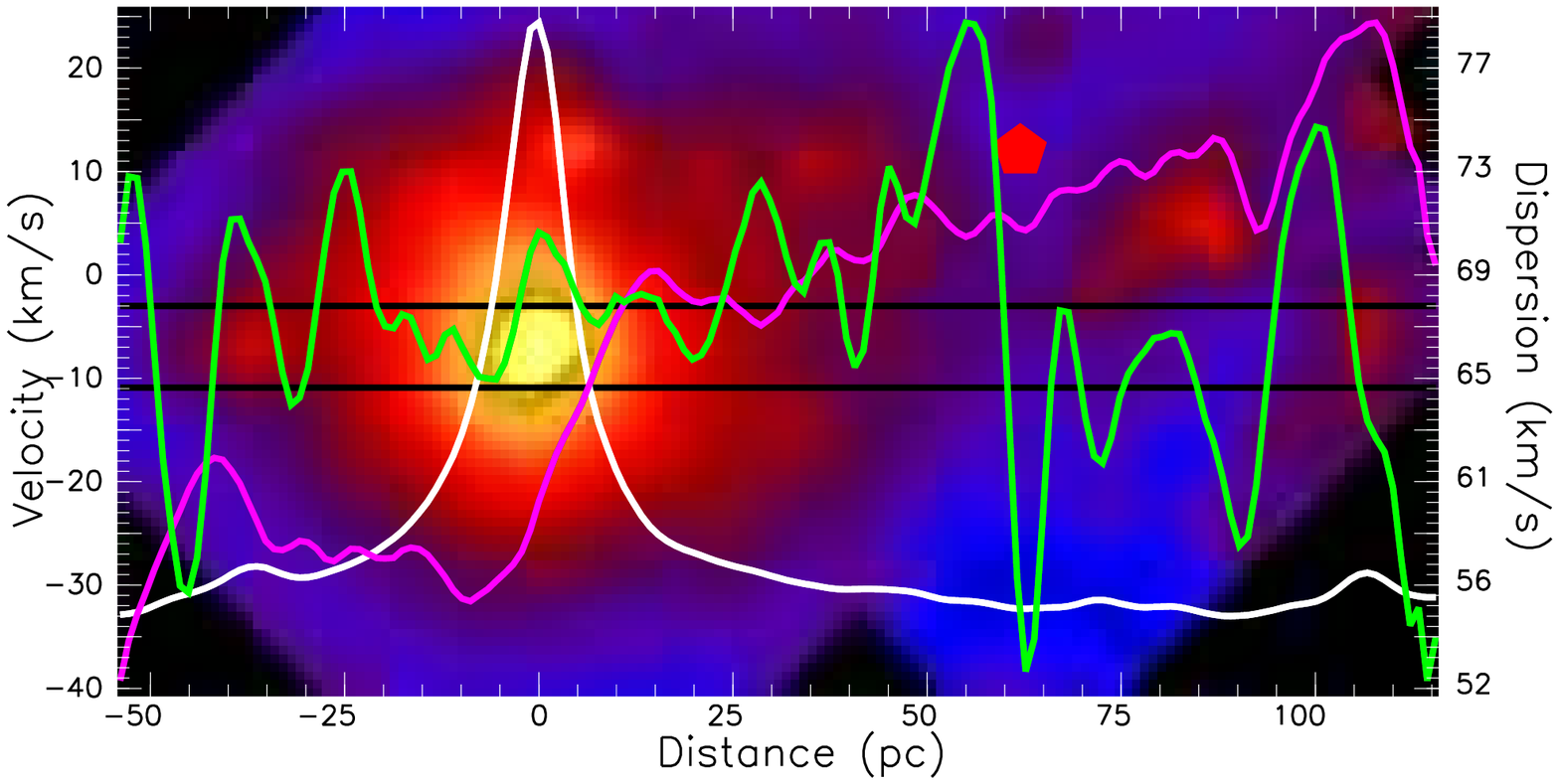}
\caption{Flux (white), velocity dispersion (green) and velocity (magenta) profiles along a pseudo-slit traced to include the optical nucleus along the direction of maximum variation of the velocity field. The location of the photometric centre of \cite{Thatte:2000p4743} is marked as a red dot. The pseudo-slit is plotted in black over the stellar continuum map for reference.}
\label{figure:slit}
\end{center}
\end{figure}

The stellar continuum derived from the first two CO absorption bandheads is mainly concentrated in the optical nucleus and, unsurprisingly, shows a similar morphology to that of the K-band image. On the other hand, the stellar kinematics show a completely different picture to that traced by the gas. Whereas the gas kinematics appear to be completely dominated by shocks and outflows at small scales, the stars show a smooth velocity gradient from northeast to southwest, typical of a rotating disk. Superimposed on this, the continuum emission from the optical nucleus is dominated by a coherent internal rotation, as highlighted in Fig.~\ref{figure:slit}. The amplitude of the projected velocity field, measured peak to peak, is $\sim32$\,km\,s$^{-1}$ within $\sim24$\,pc. The de-projected rotation velocities for an adopted inclination of $i=24^{\circ}$ would be a factor 2.5 higher. Although the uncertainties in the stellar kinematics are comparable to this value (in an spaxel basis), Fig.~\ref{figure:slit} shows a very clear jump in the projected velocity (magenta line) across the optical nucleus.

As shown in Fig.~\ref{figure:stellar}, we detect two bright spots of continuum emission (apertures H and I) in the outskirts of the optical nucleus, with a substantially high equivalent width. Given the spatial resolution of our data, these bright sources in the stellar continuum could be identified as individual stars -- late type giants or supergiants -- in a post main sequence phase. This is consistent with the main scenario proposed for the optical nucleus, discussed in Sec.~\ref{section:ON}, where the UV photons of a population of non-ionising stars would excite the molecular gas, explaining the overpopulation of the $J_{l}=3$ levels shown in Fig.~\ref{figure:h2}. Such a population of stars would be consistent with the age of the cluster derived from the CO absorption bands, as discussed in Sec.\ref{section:ON}.

\subsection{Pointing D}
\label{sec:pointingD}

\begin{figure*}[ht]
\center
\includegraphics[angle=0, , width=1.\textwidth]{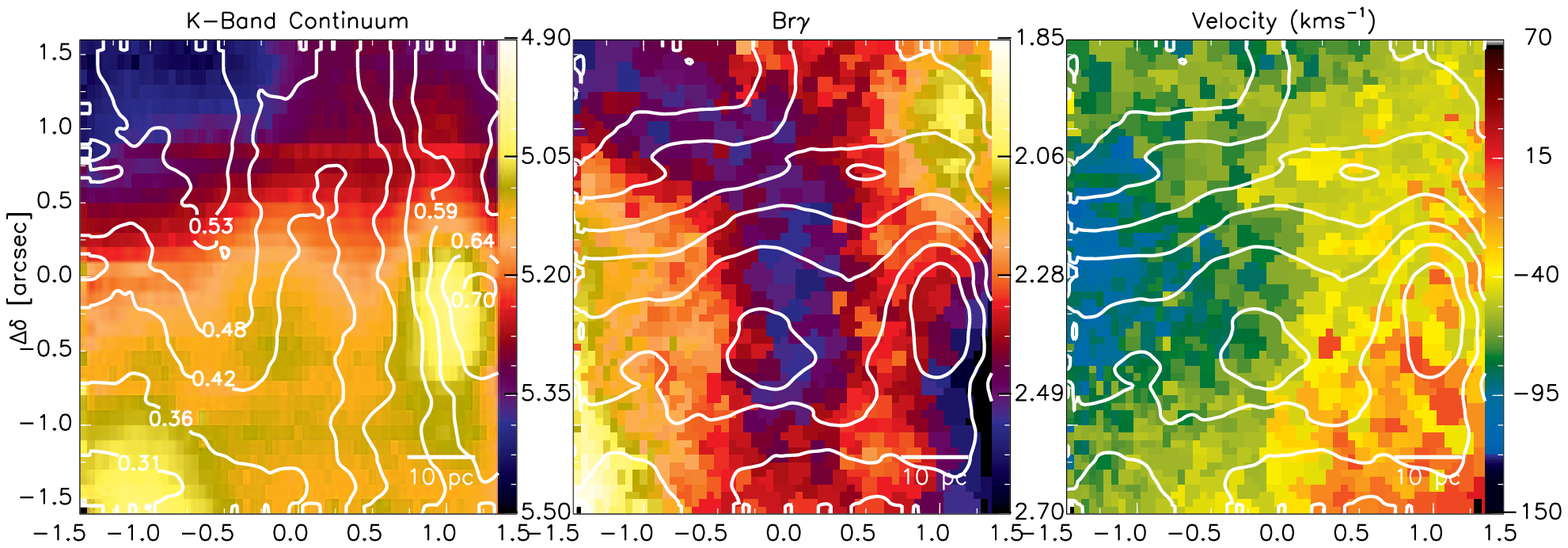}
\caption{Maps of the K-band continuum emission (left), \Brg\ surface brightness (centre) and velocity (right) from SINFONI data at pointing D. The white contours in the left panel correspond to the H-K colour obtained from the SINFONI data; the contours in the centre and right panels show the K-band continuum emission. The values of the \Brg\ velocity are in the same colour scale as those shown in Fig.~\ref{figure:maps} for the other three pointings.}
\label{figure:m83_d}
\end{figure*}

Because of the different integration time for the SINFONI data from pointing D, we have not included them in the general analysis of the emission and kinematics of the nuclear regions of M83. However, we can draw some conclusions about the hidden mass location proposed by \cite{Diaz:2006p4776}. As shown in Fig.~\ref{figure:m83_d}, we have extracted a K-band continuum image, H--K colour map and \Brg\ surface brightness and velocity maps of the $\sim65\times65$\,pc region. The H-K values obtained lie within the range $\sim0.2-0.7$, in good agreement with those derived by \cite{Wiklind:2004p5356} using NICMOS F160W and F222W images. The dust lane that crosses from north to south in the central region of M83 is just at the east of our field of view, although the extinction gradient is clearly visible towards the western part of the map. The \Brg\ maps show the north-west end of the ring feature seen in Fig.~\ref{figure:maps}, and the smooth velocity field that continues beyond pointing C. The velocity gradient does not show any evidence of a hidden mass in this position, but simply reflects the difference in velocity between two regions, the ring-like structure observed in pointing C and the bright lane of \Brg\ emission at the west of pointing D (see Fig.~\ref{figure:m83_d}).

\section{Warm molecular gas: H$_{2}$ transitions}
\label{sec:warmh2}

\begin{figure*}[t]
\begin{center}
\begin{tabular}{cc}
\includegraphics[angle=90, width=.48\textwidth]{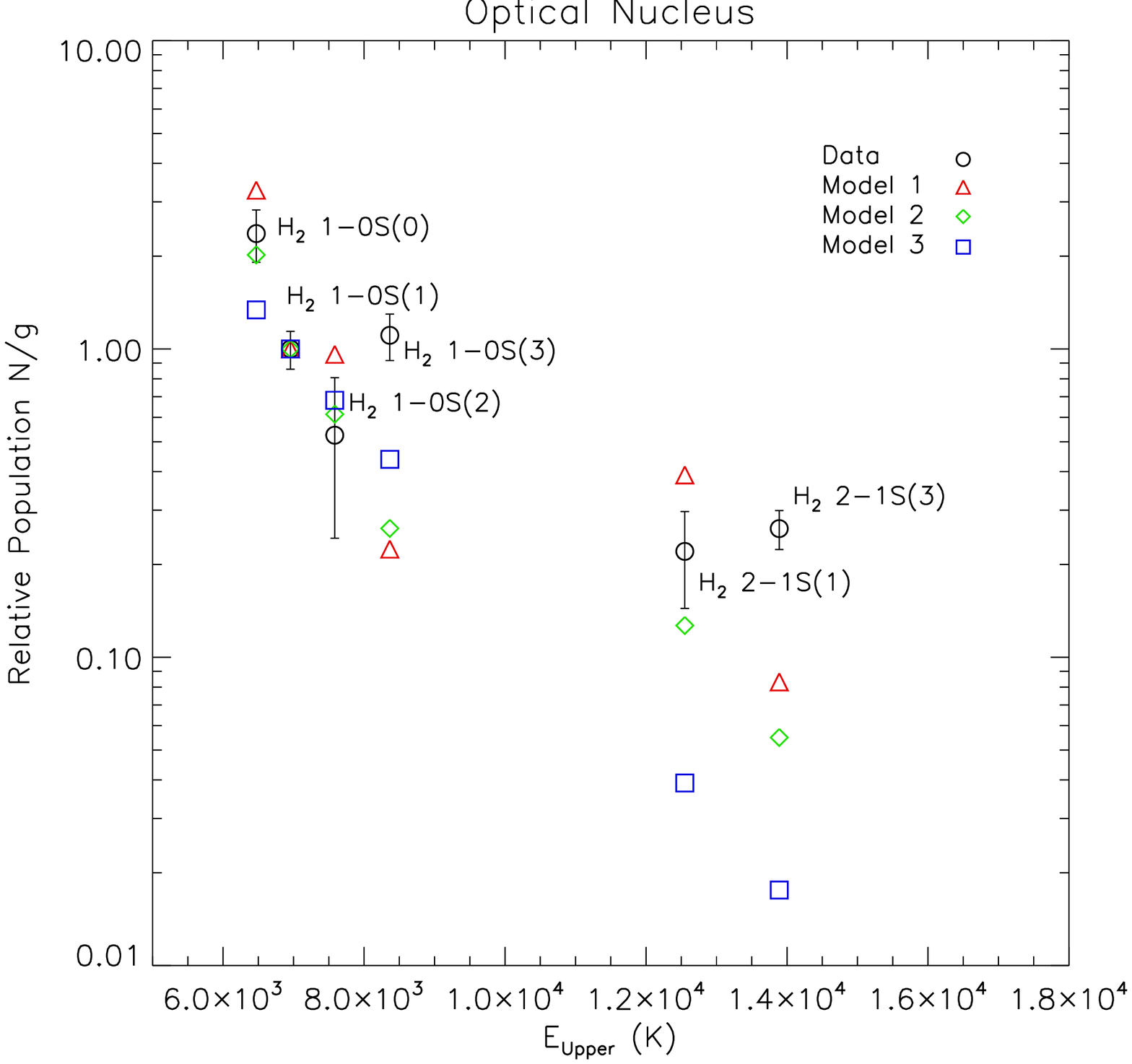} & \includegraphics[angle=90, width=.48\textwidth]{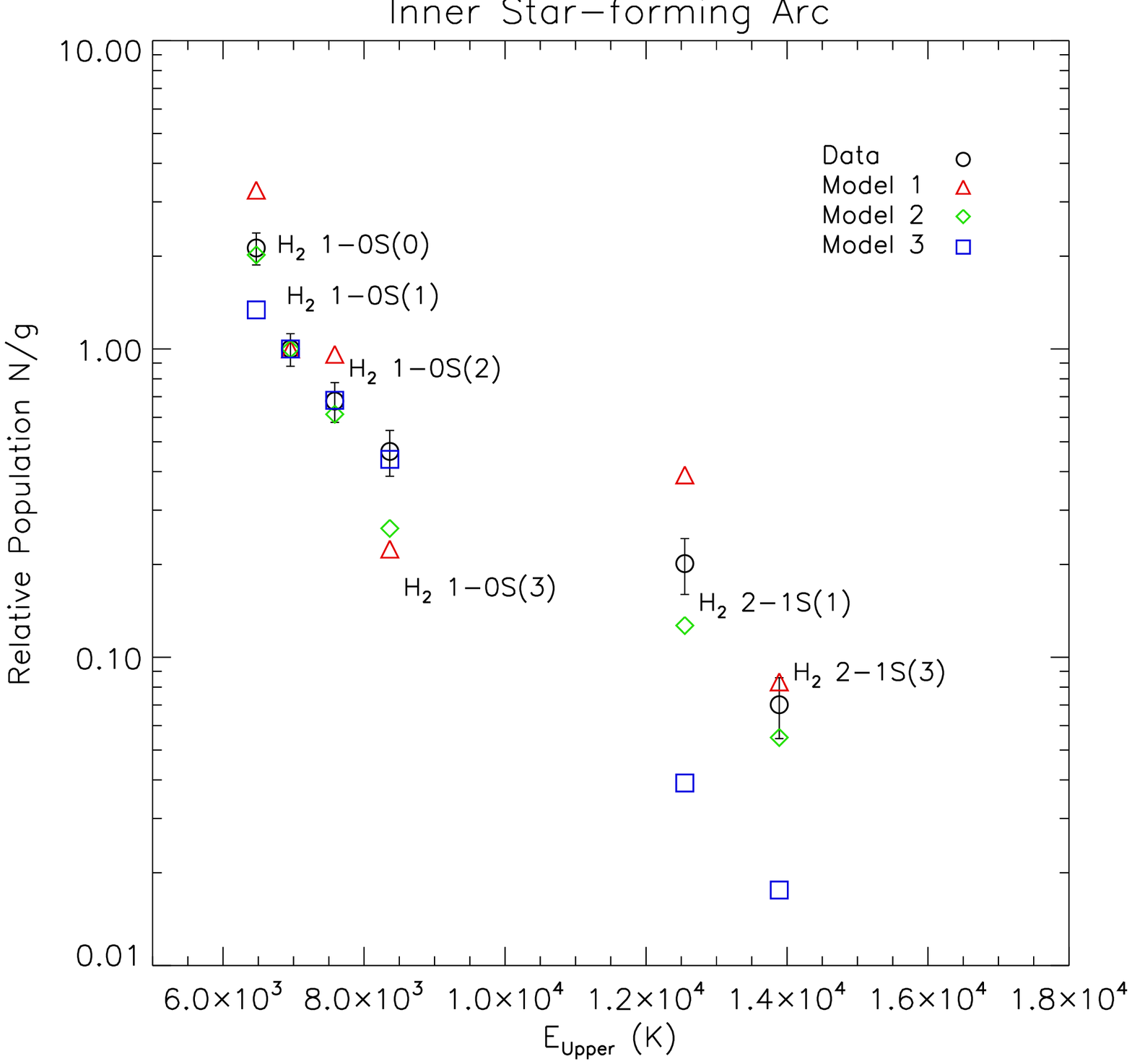} \\
\end{tabular}
\caption{H$_{2}$ excitation diagrams relative to 1-0S(1) for the optical nucleus (left panel) and the inner star-forming arc (right panel). The lines from which the population levels are derived, are indicated. Overplotted are three of the five PDR models discussed in \cite{Davies:2003p3644}. Models 1 and 2 consist of moderate and high density gas respectively where the main excitation process is UV fluorescence, while model 3 describes a fully thermalised region. As mentioned in the text, the error bars are obtained by a \emph{bootstrap} method of $N=1000$ simulations of the spectra. The diagram of the optical nucleus shows a strong overpopulation of the $J_{u}=5$ transitions that points to non-thermalized excitation mechanisms like radiative fluorescence, whereas the populations of the inner arc are consistent with a dense PDR where the $\nu=1$ levels are thermalized by collisions while the $\nu=2$ overpopulation is characteristic of fluorescent excitation.}
\label{figure:h2}
\end{center}
\end{figure*}

There are two very distinct areas of strong H$_{2}$ emission, one associated with the optical nucleus and the other with emission from the inner arc at the western part of the FoV (see Fig.~\ref{figure:maps}). We have extracted integrated spectra from both the optical nucleus and the arc, including all the spaxels from each region above a given flux threshold. This threshold has been chosen to be 15\% of the brightness of the H$_{2}$ peak, and allows us to reject most of the weakest spaxels that contribute mostly to increase the noise. We have measured the fluxes of the different transitions by fitting a Gaussian profile, and derived the uncertainties using a Monte Carlo technique. The method consists of measuring the noise in the spectra as the rms of the residuals after subtraction of the Gaussian profile. Taking into account this estimation of the noise, we construct a total of $N=1000$ simulations of our spectra where the lines are again fitted. The uncertainty of our measurements is defined as the standard deviation of the fluxes for each line. The values obtained for the fluxes of the different transitions and their uncertainties are listed in Table~\ref{table:h2}.

Using these fluxes, we can calculate the various populations of the upper H$_2$ levels associated with each transition. As shown in the population diagrams in Fig.~\ref{figure:h2}, we found that the emission on both regions has an important contribution from non-thermal processes. We have compared the population of the different levels of both regions with three of the photon-dominated region (PDR) models discussed in \cite{Davies:2003p3644}, where the excitation is dominated by far-UV photons. In Fig.~\ref{figure:h2}, model 1 and 2 consist of moderate and high density gas, $n_{\rm H}=10^3$\,cm$^{-3}$ and $n_{\rm H}=10^4$\,cm$^{-3}$ respectively, where the main excitation process is UV fluorescence, whereas model 3 describes a fully thermalised region with $n_{\rm H}=10^4$\,cm$^{-3}$ and $T=2\times10^3$\,K.

The emission from the inner arc shows that the lowest ($\nu=1$) transitions are thermalized while the higher ($\nu=2$) transitions are slightly overpopulated. This is a clear sign of fluorescent excitation mechanisms, that tend to excite the highest levels. The values are consistent with model 2 of a dense PDR in which the $\nu=1$ levels are thermalised and the $\nu=2$ levels are overpopulated by fluorescence. Intriguingly, the optical nucleus shows a stronger contribution of radiative processes, most notably in terms of an overpopulation of the $J_{\rm u}=5$ levels (equivalently stronger S(3) lines).

Although these results do not allow us to quantify the contribution from the different mechanisms, it is clear that the dominant processes are rather different in both regions: while the inner arc seems to be dominated by thermal processes compatible with episodes of recent supernova activity, the contribution from radiative processes in the optical nucleus associated to fluorescent excitation mechanisms is highly significant.


\begin{deluxetable}{cccc}
\tablecolumns{4}
\tablewidth{0pt}
\tablecaption{Integrated fluxes of the H$_2$ lines for the optical nucleus and the inner arc.\label{table:h2}}
\tablehead{
\colhead{Line} & \colhead{$\lambda$} & \multicolumn{2}{c}{Flux}\\
& \colhead{($\mu$m)} & \multicolumn{2}{c}{(10$^{-15}$ erg\,s$^{-1}$\,cm$^{-2}$)} \\
&& \colhead{Optical Nucleus} & \colhead{Inner Arc}
}
\startdata
1-0S(3) & 1.958 &$4.24\pm0.73$&$1.19\pm0.20$\\
1-0S(2) & 2.034 &$0.50\pm0.27$&$0.43\pm0.06$\\
2-1S(3) & 2.073 &$1.30\pm0.19$&$0.23\pm0.05$\\
1-0S(1) & 2.122 &$1.85\pm0.26$&$1.24\pm0.15$\\
1-0S(0) & 2.223 &$0.73\pm0.14$&$0.44\pm0.05$\\
2-1S(1) & 2.248 &$0.55\pm0.19$&$0.34\pm0.07$\\
\enddata
\end{deluxetable}

\section{Emission and Kinematics of the Gas: Evidence for Supernovae}
\label{sec:supernovae}

As shown in Fig.~\ref{figure:maps}, the kinematics of the gas is totally unrelated to the stellar kinematics (Fig.~\ref{figure:stellar}). There is no clear evidence of a single uniform velocity gradient, as seen for the stars, in any of the emission line maps. Together with the dispersion, this suggest that the gas kinematics are at small scales dominated by shocks and flows. In this section we argue that these characteristics are related to the presence of supernovae.

\begin{figure}[t]
\begin{center}
\includegraphics[angle=180, width=0.48\textwidth]{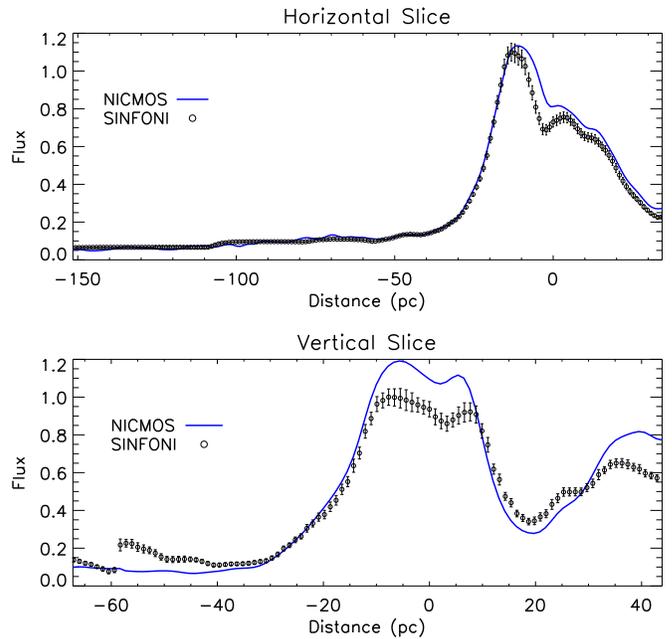}
\caption{Flux profiles in arbitrary units of the \Pa\ ring-like emission from NICMOS and SINFONI. Distances are measured from the centre of the ring. Both images were previously PSF-matched and normalised to the background of the emission before comparing the data. There is a significative relative decrease of $\sim$6\% in flux between the two epochs.}
\label{figure:profiles}
\end{center}
\end{figure}

The nature of the \Brg\ ring-like feature noted in Sec.~\ref{sec:overview} could be explained in terms of a light echo from a recent type II supernova explosion. This would be consistent with the low velocity dispersion and the presence of the complete Brackett series in the spectrum (see the spectrum of aperture A in Fig.~\ref{figure:spectra}). The projected radius of the ring is $\sim23$\,pc, which means that, if its origin was a supernova event, the explosion would have occurred $\sim75$\,yr ago. If it is indeed a light echo, one might expect to see changes on a timescale of 10 years, for example between the NICMOS data from 1998 and our data from 2009. We have compared the \Pa\ luminosity profiles of the ring measured by NICMOS with our data to confirm a possible evolution with time. We first matched the PSFs, and then extracted horizontal and vertical profiles centred in the centre of symmetry of the ring feature in both images. We applied a single scaling and normalisation to the profiles, derived to match the background of the emission far from the  \Pa\ ring. The two resulting profiles of the \Pa\ emission are shown in Fig.~\ref{figure:profiles}. Considering the expansion rate of a light echo, the difference in size of the ring between the two datasets is expected to be of $\sim$3.4\,pc, which is less than a resolution element in our coarse sampling (see Table~\ref{table:PSF}). However, there is a significant relative decrease of $\sim$6\% in brightness between the two epochs that cannot be explain as a PSF or AO effect. This decrease in the emission supports the hypothesis of a transient event like a supernova explosion. 

In Sec.~\ref{sec:overview} we also pointed out two bright [FeII] spots located at the outskirts of the optical nucleus, labeled as E and F in Fig.\ref{figure:maps}, which are associated with high velocity dispersion and are probably tracing individual supernovae. The spectrum of one of these sources (E) is shown in Fig.~\ref{figure:spectra}. The region labeled as D in Fig.~\ref{figure:maps} corresponds to the source M83-SNR-N-01 identified in \cite{Dopita:2010p5365} in their Table~3, and shows some characteristics expected for a recent supernova, i.e. strong [FeII] and H$_{2}$ emission.

There are also a few other supernova remnant (SNR) candidates that lie within our FoV, but none of them are obviously detected. The reason why there is no sign of most of these sources is because the selection criteria of the SNR candidates in \cite{Dopita:2010p5365} is in terms of the [OII] emission at 3727\,\AA\ and 3729\,\AA. To achieve high [OII]/H$\alpha$ ratios a radiative shock of $\sim$300--500\,kms$^{-1}$ is needed (\citealt{Dopita:1995p6050}, \citeyear{Dopita:1996p6053}), so it is the ionized pre-shock that actually emits in [OII]. Therefore, the temperature and the shock speed are too high to expect [FeII] or H$_{2}$ emission (see \citealt{Burton:1990p5054}). We have also assessed the list of individual X-ray sources of \cite{Soria:2002p5354} but only three of them lie within our FoV (sources 37, 40 and 43). Source 43 corresponds to the optical nucleus, source 37 is shifted by $\sim$0.7\arcsec\ to the west and by $\sim$0.5\arcsec\ to the south of our aperture A and source 40 is shifted $\sim$0.5\arcsec\ to the west and $\sim$0.6\arcsec\ to the south of aperture B. Taking into account the uncertainties in the position of the X-ray sources, source 37 could be tentatively associated with the ring of \Brg\ emission.

As proposed by \cite{Raymond:2001p6005}, the width of the [FeII] and H$_{2}$ lines is a good estimation for the velocity of the shock speed in SNRs. The [FeII] line is clearly broadened with an intrinsic width of $\sim$100--180\,kms$^{-1}$. In contrast, the H$_{2}$ line width is no more than $\sim$40\,kms$^{-1}$, indicating it is barely resolved. These values are in good agreement with the values derived in \cite{Burton:1990p5054} for fast J shocks (i.e. ``jump" shocks, exhibiting discontinuous transition; in contrast to ``continuous" shocks showing continuous transition in velocity, density and temperature) where the [FeII] emission is expected to be stronger, and slow J or fast C shocks for the molecular hydrogen emission. Given the size of these regions of $\sim$4\,pc or less and the velocity dispersion of $\sim$100--180\,kms$^{-1}$, and assuming a constant expansion rate, we obtain an upper limit to their age of $\sim4\times10^{4}$\,yr which supports the argument that they are recent events.

The global velocity gradient in the [FeII] emission of $\sim$60\,kms$^{-1}$ along the arc is very similar to that observed for the \Brg\ line. However, the extended [FeII] emission throughout the inner arc and the high velocity dispersion are rather different compared to the \Brg\ and H$_{2}$. Both are consistent with a scenario where the most massive stars have already exploded as supernovae. These explosions would blow up the gas of a thin quiescent disk in the perpendicular direction.

This would provide a natural explanation for the systematically higher velocity dispersion observed for the [FeII] emission along the inner arc: the shocks of the supernovae would be fast enough to dissociate the H$_{2}$ molecules and enhance the gas-phase Fe abundance as well as to generate singly ionised Fe, but not to fully ionise the H.

A scenario in which many stars have recently exploded as supernovae, or may soon do so, is supported by the strong radio continuum found by \cite{Saikia:1994p5355} and the diffuse X-ray emission detected in the arc (see \citealt{Soria:2002p5354}). These authors also found a high abundance of Ne, Mg, Si and S with respect to Fe. This suggest that the interstellar medium would have been enriched due to type-II supernova explosions in the recent past. This scenario would also provide an origin for the radiative fluorescence excitation of the H$_{2}$ molecules. It can be explained in terms of a population of stars of $\sim$5--8\,M$_{\odot}$ that have not exploded yet as supernovae (and may not do so), but emit enough UV radiation to excite the H$_{2}$ \citep{Puxley:1990p5361}.

This scenario would also be in good agreement with the ages of optically-selected star clusters inferred in \cite{Harris:2001p5658}. Based on HST WFPC2 observations, these authors studied the star formation history of the southern star-forming arc and found that more than $\sim$75\,\% of the more massive clusters (M$\gsim2\times10^4\,M_{\odot}$) have ages less than 10\,Myr. They found a sharp cutoff in the age distribution of the star clusters, and proposed a recent burst of star formation activity that began $\lsim$10\,Myr ago, suggesting an outward propagation across the arc.

\begin{figure*}[t]
\begin{center}
\includegraphics[angle=90., , width=1.\textwidth]{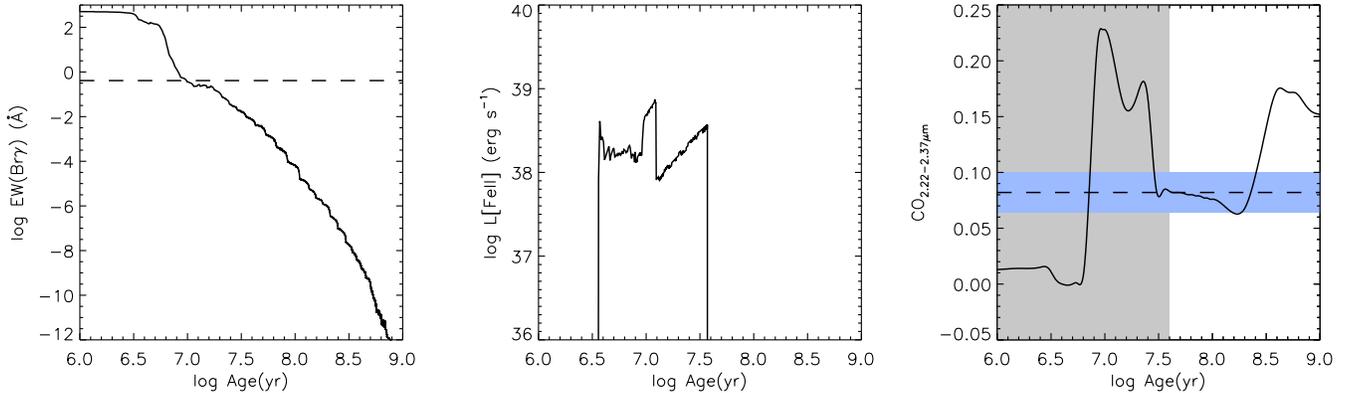}
\caption{Evolution of the \Brg\ equivalent width (left) and [FeII] flux (centre) from SB99 models, and CO$_ {\rm 2.22-2.37\mu m}$ index (right) from M05 models. The dotted line in the left panel shows the upper limit of the \Brg\ equivalent width at the optical nucleus. We have assumed the theoretical ratio from \cite{Colina:1993p6004} to convert the [FeII] 1.26\,$\mu$m flux predictions to [FeII] 1.64\,$\mu$m fluxes. The upper limit for the [FeII] luminosity at the optical nucleus is $\log \rm[FeII] (\rm erg\,s^{-1}) \lsim 33$, far below the expected luminosity of a $\sim10$\,Myr population. In the right panel, the region incompatible with the low \Brg\ and [FeII] fluxes is shown in grey. The blue strip shows the range of CO$_ {\rm 2.22-2.37\mu m}$ index allowed by our data to 1$\sigma$ confidence.}
\label{figure:maraston}
\end{center}
\end{figure*}

We can make a rough estimate of the age of the youngest stellar population along the inner star-forming arc under the assumption that the [FeII] emission originates in supernovae. The STARBURST99 stellar population synthesis models (\citealt{Leitherer:1999p6938}, SB99 hereafter) predict that this emission would reach its maximum at $\sim$10\,Myr (see Fig.~\ref{figure:maraston}). On the other hand, the \Brg\ emission also appears to be dominated by a supernova event. It shows little ionized gas emission from the star themselves and it is probably part of the tail end of the supernovae for the same burst of star formation that will be discussed in Sec.~\ref{section:ON}. This implies that there are few OB stars left and that the star formation episode was consequently a short burst. Compared to the arc, the optical nucleus shows more H$_{2}$ emission and less [FeII] while still having very little \Brg\ emission. This is also suggestive of a short burst of star formation. Both regions therefore appear to have experienced an episode of star formation around $\gtrsim$10\,Myr ago.
More detailed assessments of the age of the star formation in the optical nucleus, based on dynamics as well as spectral features, are presented in Sec.~\ref{section:ON}.

\section{Optical Nucleus: An Evolved Massive Off-Nuclear Star Cluster}
\label{section:ON}

In order to shed some light on what one should consider to be the nucleus of M83, we have studied in more detail the kinematic and photometric properties of the optical nucleus. We made use of NIC2 F222M K-band continuum image to fit the brightness profile of the cluster. We adopted a model of a symmetric S\'ersic profile convolved with the PSF to fit the core of the nucleus, combined with an asymmetric Gaussian to take care of the extended emission. The fits reveal a strongly peaked nucleus of $R_{\rm eff}=2.97\pm0.15$\,pc (with a S\'ersic index of $n=2.7$), clearly unresolved in the SINFONI data, which sits on a more extended emission of size $\sim$50\,pc. Given that the core of the cluster appears point-like and is not resolved in the AO data, we use this value as an upper limit for the effective radius of the cluster.

\subsection{Age and Mass Estimation from Stellar Kinematics}
\label{sec:age_mass_1}

We have extracted an integrated spectrum of the core of the nucleus using an aperture of radius $R_{\rm eff}$, and performed a similar analysis as for the stellar kinematics, making use of the pPXF code to fit it with a library of stellar templates. The results of the fitting yield a velocity dispersion of $\sigma_{\rm R_{\rm eff}} \simeq 71 \pm 14$\,km\,s$^{-1}$. We have adopted this value to provide an estimate of the dynamical mass of the cluster, given by the following relation:

\begin{equation}
M_{\rm dyn} = \eta {\sigma^2 R_{\rm eff} \over G}
\end{equation}

where $\eta$ is a geometric constant, $\sigma$ is the velocity dispersion, $R_{\rm eff}$ is the effective radius of the cluster and $G$ is the gravitational constant. The geometrical constant $\eta$ is determined by the density and velocity distribution of the cluster, and is commonly assumed to be $\sim$10 when $R_{\rm eff}$ is used \citep{Walcher:2005p6759, McCrady:2007p6226} to obtain the total mass of the cluster. However, different values from 3 to 10 could be adopted (see \citealt{Hagele:2009p6754}, \citealt{Barth:2009p6749} and \citealt{BasuZych:2009p6282}). As shown in Fig.~\ref{figure:slit}, the optical nucleus has its own internal velocity gradient, distinct from the main velocity gradient of the galaxy, that suggests that the cluster is virialized. Using the lowest value of $\eta = 3$ that assumes an isotropic velocity field, yields $M_{\rm dyn} = (1.1\pm0.4)\times10^7\,M_{\odot}$ for the mass of the cluster, a value consistent with an earlier stellar kinematic estimate by \cite{Thatte:2000p4743}.

Using both NICMOS F222M image and our SINFONI data, we have obtained the K band luminosity of the star cluster. In order to perform synthetic photometry in our IFU data, we made use of the K band response curve of 2MASS as defined in \cite{Cohen:2003p3372} to obtain a synthetic K-band image.  We have used the same aperture of radius 2.97$\pm$0.15\,pc to measure the luminosity of the star cluster in both images. The values obtained are $L_{\rm K, SINFONI}=6.03\times10^5\,L_{\odot}$\footnote{$L_{\odot}=3.826\times10^{26}$\,W.} and $L_{\rm K, NICMOS}=5.43\times10^5\,L_{\odot}$ from the SINFONI and NICMOS data respectively.

We can thus derive a mass-to-light ratio $M_{dyn}/L$ ratio based on the dynamical mass obtained above, which is $M_{dyn}/L\sim0.29\pm0.12\ M_\odot/L_{K,\odot}$ (where the denominator is in units of monochromatic solar K-band luminosity $L_{\rm K,\odot}=2.150\times10^{25}$\,W). In order to make a first estimation of the age of the cluster, we have used the stellar population synthesis models of \citet[][M05 hereafter]{Maraston:2005p2885}. The age that corresponds to the mass-to-light ratio derived above according to these models is $\log T(\rm yr) = 8.17^{+0.08}_{-0.36}$.

\subsection{Age and Mass Estimation From Spectral Diagnostics}
\label{sec:age_mass_2}

In addition to the age derived from stellar kinematics and photometry, we can put further constrains on the age via the CO stellar absorption. We have measured the equivalent width of the first CO bandhead as well as its CO index (defined as the ratio of the flux densities at 2.37$\mu$m and 2.22$\mu$m) to compare the values with the prediction from the SB99 and the M05 synthesis models respectively. The value obtained for the CO equivalent width is $W_{\rm CO}=11.06\pm0.25$\,\AA, where uncertainties are calculated by a Monte Carlo method of $N=1000$ simulations of the spectrum. 

To compare the equivalent width with the SB99 models (assuming a Salpeter IMF and solar metallicity), we have considered an instantaneous burst of star formation as implied by the low \Brg\ emission from the cluster. It is compatible both with a young population of less than $\log T(\rm yr)\sim7.17$ or with an evolved population older than $\log T(\rm yr)\sim7.70$. On the other hand, according to the models, the [FeII] emission suggests a lower limit of $\log T(\rm yr)\sim7.55$ for the age of the star cluster. Combining the constraints from the [FeII] emission and the $W_{\rm CO}$ yield an estimate for the age of the cluster of $\log T(\rm yr) = 7.97^{+0.12}_{-0.33}$. 

In contrast to SB99, the M05 models use the CO index. The integrated spectrum of the optical nucleus yields CO$_ {\rm 2.22-2.37\mu m} = 0.082\pm0.018$. Using the M05 models with the same constraints as discussed above from the \Brg\ and [FeII] emission, the resulting age is 
$\log T(\rm yr) = 7.98^{+0.43}_{-0.37}$.

It is notable that the ages estimated here from the CO equivalent width and index are very similar to that calculated previously from the mass-to-light ratio using the dynamical mass. Instead of comparing ages derived using the different methods, we can instead compare masses. To do this, we first estimate the mass-to-light ratio $M^{\rm \star}/L$ from the M05 model associated with the age derived from the CO index. We can then derive the stellar (rather than dynamical) mass for the core of the star cluster. Again adopting a Salpeter IMF and solar metallicity, we find $M^{\rm \star}/L\simeq 0.22\pm0.08$. Since the luminosity is measured within $R_{\rm eff}$ (which by definition contains half the light), we set the luminosity of the cluster to be twice that given in Sec.~\ref{sec:age_mass_1}. Thus the stellar mass is $M^{\rm \star} \simeq (7.8\pm2.4)\times10^6\,M_{\odot}$.

This value is a factor of a few higher than the photometric mass estimate from \cite{Thatte:2000p4743} and rather similar to that given in \cite{Wiklind:2004p5356}. We note that, within the inevitable uncertainties, it is also remarkably consistent with the dynamical mass we estimated previously.

These ages (from \ref{sec:age_mass_1} and \ref{sec:age_mass_2}) are a little higher, but not inconsistent with that discussed at the end of Sec.~\ref{sec:supernovae} and all point to an age approaching, but perhaps a little less than, $\sim100$\,Myr. We have considered only instantaneous star formation, since continuous models are clearly ruled out by the low \Brg\ equivalent width. However, a short but finite burst length of, for example, $\sim50$\,Myr could plausibly reconcile the small differences between the estimates.

\subsection{Could the Optical Nucleus Host a Supermassive Black Hole?}

As a particularly massive star cluster, and potentially the nucleus of M83, the optical nucleus is a suitable candidate for hosting a supermassive black hole. We have therefore estimated the mass of the supermassive black hole that one might expect to find in the inner regions of M83. Using GALFIT \citep{Peng:2010p6520}, we fitted a three component model to the 2MASS K-band image \citep{Skrutskie:2006p5780}, that allows us to separate the contribution of the disc, bar and bulge to the total flux distribution. The K-band luminosity of the bulge derived from the fit is log$L_{\rm K,bulge}=9.705$ in $L_{\rm K,\odot}$ units. Taking into account the $M_{\rm BH}/L_{\rm K,bulge}$ relation from \cite{Marconi:2003p6211}, we estimate the mass of the central black hole should be around $\sim3.9\times10^6\,M_{\odot}$. This value for the BH mass is also similar to that obtained from the M$_{\rm BH}$ -- $\sigma$ relation \citep{Tremaine:2002p7071}, if we take 100\,kms$^{-1}$ from Fig.~\ref{figure:stellar} to be the velocity dispersion of the stars in the bulge. Although this mass is less than the dynamical mass of the optical nucleus, it seems unlikely that a supermassive black hole should make up more than $\sim35$\% of the dynamical mass of a star cluster.

On the other hand, the measurements from \emph{Chandra} presented in \cite{Soria:2002p5354} show that the optical nucleus is one of the brightest sources in X-rays within the nuclear regions of M83. They fit the nuclear spectra to a power-law model with total X-ray luminosity (0.3--8\,keV) of $L_{\rm X}\sim2.6\times10^{38}$\,erg\,s$^{-1}$ and a photon index of $\Gamma\sim$1.15 (see their Table~3). The total luminosity is compatible with the X-ray emission from either a low luminosity AGN or a stellar-mass black hole candidate in a binary system \citep{Grimm:2003p8619}. However, the lack of any other AGN signature (e.g. [SiVI] emission at 1.96\,$\mu$m) indicate that if a supermassive black hole were present, it would not be in an AGN phase at the present epoch. It is likely then that the X-ray emission comes from X-ray binaries in the hard state.

\section{Location of the Nucleus}
\label{section:location_nucleus}

The location of the nucleus of M83 has been topic of an intense debate over the last decade. \cite{Thatte:2000p4743} first reported the discovery of a double nucleus in M83 based on long-slit measurements of the velocity dispersion of stars. They found two peaks in the velocity dispersion, one coincident with the optical nucleus, and another associated with the centre of symmetry of the bulge isophotes, both enclosing a dynamical mass of $\sim1.3\times10^{7}\,M_{\odot}$. However, the lack of two-dimensional information did not allow them to obtain the precise position of this second mass concentration.

\cite{Mast:2006p4849} made optical integral field spectroscopic observations of the inner 12\arcsec $\times$ 21\arcsec\ of the galaxy that allow them to study in more detail the velocity field of the ionized gas. Their results support the conclusion about the presence of a second mass concentration of $\sim1.0\times10^{7}\,M_{\odot}$ (although it is possible that its position was misplaced due to an incorrect spatial scaling).

\cite{Diaz:2006p4776} pointed out the position of a mass concentration at the northernmost part of the southern star-forming arc. They estimated that this hidden concentration would enclose a total mass of $\sim1.6\times10^{7}\,M_{\odot}$, derived from the ionized gas kinematics. They also found that the position of this hidden mass corresponds to a peak of emission in the mid-IR continuum at 10\,$\mu$m.

More recently, \cite{Houghton:2008p4894} combined near-IR long-slit spectroscopy with HST imaging to study the 20\arcsec $\times$ 20\arcsec\ central region of M83. They looked at the stellar kinematics for dynamical signatures of putative hidden mass concentrations at locations indicated in previous work. Their results show no evidence of obscured masses and they conclude that the velocity gradients observed in the gas kinematics are a consequence of shocks.

Making use of near-IR integral field spectroscopy and numerical simulations, \cite{Rodrigues:2009p5531} presented a detailed study of the ionised gas kinematics in the inner 5\arcsec $\times$ 13\arcsec\ of the galaxy, covering the wavelength interval from 1.2\,$\mu$m to 1.4\,$\mu$m with a spatial sampling of 0\farcs36. They focused on the dynamical properties and evolution of the optical nucleus, the CO kinematic centre (similar to the photometric centre) and the putative mass concentration proposed by \cite{Diaz:2006p4776} that is coincident with the 10$\mu$m continuum emission peak. Based on the ionised gas kinematics, they derived dynamical masses of $\sim6.0\times10^{7}\,M_{\odot}$, $\sim4\times10^{6}\,M_{\odot}$ and  $\sim2.0\times10^{7}\,M_{\odot}$ for the kinematic centre, the optical nucleus and the hidden mass concentration respectively.

Finally, \cite{Knapen:2010p5390} analyzed in detail the ionized gas kinematics, making use of Pa$\beta$ IFU observations with a spatial sampling of 0\farcs36, confirming the results found by \cite{Thatte:2000p4743} that the photometric centre coincides with the kinematic centre, and this location is offset by $\sim$4\arcsec\ ($\sim$90\,pc) from the optical nucleus. They proposed two possible options for the location of the true nucleus of M83.  One option is the presence of an obscured hidden mass in the kinematic and photometric centre, that would require a dust extinction of $A_{V}=3-10$\,mag. However, authors consider this option unlikely, because no other signatures -- such as a peak in the velocity dispersion or in the near-IR emission -- of a hidden mass are found. Their other option was that the optical nucleus is the true nucleus, and it is displaced from the kinematic centre as a result of some past interaction.

Our results clearly show that on scales of tens of parsecs, the gas kinematics are dominated by shocks and outflows. But we have also seen there is a global gradient of $\sim$60\,kms$^{-1}$ along the inner arc which is totally unrelated to the stellar velocity field. Given the orientation of the galaxy, these gas kinematics appear to be tracing an inflow of gas along the inner bar to the photometric centre. We would expect such a bar-driven gas inflow to terminate at the nucleus, and the location of the end of the inflow does coincide with the photometric centre.
We also emphasize that there is an increase of the stellar continuum at this point, consistent with the centre of the bulge. But this continuum, primarily from older stars, is mostly swamped by the stronger continuum from the young stars in the surrounding star-forming ring. The lack of a very recent starburst in the photometric centre would explain the low stellar continuum emission in this region.

As mentioned before, we have not found any evidence or signature of a hidden mass near the kinematic centre. However, the velocity dispersion expected due to random motions around such a black hole, given the spatial resolution of our AO data, would be $\sigma_{\rm BH}\sim60$\,kms$^{-1}$. Taking into account the instrumental resolution of our data ($R\sim1500$, or an instrumental broadening of $\sigma_{\rm instr}\sim85$\,kms$^{-1}$) and the dispersion of the stars of $\sigma_{\rm \star}\sim100$\,kms$^{-1}$ measured almost everywhere, the effective enhancement of the dispersion expected to be observed for this BH would be of $\sim$10\,kms$^{-1}$, making our data insensitive to such a compact object. Thus, the lack of a kinematic signature of a supermassive BH does not equate to the absence of such a BH at the location of the kinematic and photometric centre of M83.

We also considered the possibility that the $\sim3.4$\arcsec\ offset of optical nucleus from the kinematic centre could be the result of an $m=1$ perturbation in the gravitational potential, as suggested by \cite{Knapen:2010p5390}. As shown in \cite{Bournaud:2005p6704}, an $m=1$ perturbation of the potential could be explained by asymmetric accretion of gas towards the inner regions of the galaxy, which would be in agreement with the unperturbed spiral pattern revealed at mid-infrared wavelengths for M83 \citep{Dale:2009p6943}. Alternatively, as discussed in \cite{Hopkins:2010p6217}, an eccentric pattern resulting from a past interaction could persist up to $\sim10^4$ dynamical times. However, a model of an eccentric disk as used by \cite{Tremaine:1995p6216} to explain the double nucleus of M31 is probably not appropriate to explain the offset in here. It is caused by the high density of stars near apocenter in their elliptical orbits: the combination of their slow velocities at this point together with the fact that their velocity is along the line of sight, create the illusion of a secondary nucleus. In M83, the compactness of the optical nucleus and the presence of a coherent internal velocity gradient (see Fig.~\ref{figure:slit}) are inconsistent with the rather diffuse appearance expected for such a gravitational perturbation. Furthermore, the presence of an apparent secondary nuclei as in M31 is favoured by the close edge-on sightline. In principle, the eccentric disk could obscure the radiation from the black hole, even at X-ray wavelengths, and would explain the lack of signature of a compact object near the kinematic centre of M83. However, the low inclination of the galaxy and the fact that the optical nucleus lies within the rather symmetric circumnuclear ring traced by the molecular gas (see \citealt{Sakamoto:2004p5705}) make it unlikely.

\section{Conclusions}

We have presented new and detailed near infrared adaptive optics integral field spectroscopic data for the innermost $\sim$200\,pc of M83, and studied the kinematics and distribution of the stars as well as molecular and ionised gas. Our conclusions are as follows:

\begin{enumerate}

\item
The stellar kinematics show a smooth global velocity field typical of uniform rotation. They reveal an independent and coherent velocity gradient intrinsic to the optical nucleus with an amplitude of $\sim$30\,kms$^{-1}$. The velocity dispersion across the whole region also shows a smooth distribution, with values in the range $\sim50$--$100$\,km\,s$^{-1}$.

\item 

The ionized and molecular gas reveal a complex situation in which their kinematics are completely dominated by shocks and inflows at small scales but trace globally an inflow along the nuclear bar to the kinematic centre. This, and the fact that they are totally unrelated to the stellar kinematics, make them unsuitable to estimate dynamical properties of the central regions. 

\item
There is plentiful evidence for recent supernovae. A bright ring-like \Brg\ feature, which dominates the inner star-forming arc, has low dispersion and no measurable velocity gradient. It can be explained in terms of a light echo from a recent supernova explosion. The [FeII] emission has high dispersion along the arc suggestive of shocks from supernova remnants; and in the optical nucleus shows two bright locations where the dispersion is high, which are likely to be individual supernova remnants. A comparison of the gas and stellar kinematics indicates that the off-nuclear mass concentrations, which had been proposed on the basis of ionised gas kinematics, are instead regions where there are complex kinematics associated with recent supernova events.

\item
A spatial study of the excitation mechanisms of the warm H$_{2}$ suggest that the inner arc is dominated by collisional mechanisms, consistent with an episode of recent supernova events, while H$_2$ in the optical nucleus has a higher contribution from radiative processes.

\item
The $\sim3$\,pc effective radius of the optical nucleus together with its coherent internal kinematics yield a dynamical mass of $M_{\rm dyn} = (1.1\pm0.4)\times10^7\,M_{\odot}$. Its K-band luminosity is $L_{\rm K}=5.7\times10^5\,L_{\odot}$.
The resulting mass-to-light ratio implies an age of $\sim100$\,Myr that is fully consistent with that implied independently by the CO index and equivalent width. Similarly, the age implied by the CO index yields a stellar mass for the cluster of $M^{\rm \star} \simeq (7.8\pm2.4)\times10^6\,M_{\odot}$, consistent with the dynamical mass.

\item
We show that the optical nucleus cannot be an $m=1$ perturbation, and is not the `true' nucleus of M83. Instead, we argue the this is indeed located at the photometric and kinematic centre of M83's bulge, where there is a measurable peak in the K-band continuum (albeit swamped by the bright surrounding star forming ring). We also show that, for the expected black hole mass in the centre of M83, one would not expect to see a dynamical signature at currently attainable spatial resolutions.
\end{enumerate}

\acknowledgments
We thank the anonymous referee for useful comments and suggestions. JPL and LC acknowledge support by the Spanish Plan Nacional del Espacio under grants ESP2007-65475-C02-01 and AYA2010-21161-C02-01. Part of this research was perform while JPL benefited from a short-term stay at the Max Planck Institute for Extraterrestrial Physics (MPE) under grant BES-2008-007516.

JPL wants to thank the support and hospitality of the MPE Infrared and Submillimeter Astronomy Group. JPL also wants to thank Miguel Pereira-Santaella, Daniel Miralles, Ruyman Azzollini  
and Hauke Engel
for fruitful discussions.

This research is based on observations collected at the European Southern Observatory. We acknowledge use of data products from the Two Micron All Sky Survey, which is a joint project of the University of Massachusetts and the Infrared Processing and Analysis Center/California Institute of Technology, funded by the National Aeronautics and Space Administration and the National Science Foundation.

Some of the data presented in this paper were obtained from the Multimission Archive at the Space Telescope Science Institute (MAST). STScI is operated by the Association of Universities for Research in Astronomy, Inc., under NASA contract NAS5-26555. Support for MAST for non-HST data is provided by the NASA Office of Space Science via grant NNX09AF08G and by other grants and contracts.

\bibliographystyle{apj}
\bibliography{apj-jour,aa}
\end{document}